\documentclass[aps,pra,twocolumn,10pt,superscriptaddress,showpacs]{revtex4-1}

\def\bra#1{\mathinner{\langle{#1}|}}
\def\ket#1{\mathinner{|{#1}\rangle}}
\def\braket#1{\mathinner{\langle{#1}\rangle}}

\let\protect\relax
{\catcode`\|=\active
  \xdef\Braket{\protect\expandafter\noexpand\csname Braket \endcsname}
  \expandafter\gdef\csname Braket \endcsname#1{\begingroup
     \ifx\SavedDoubleVert\relax
       \let\SavedDoubleVert\|\let\|\BraDoubleVert
     \fi
     \mathcode`\|32768\let|\BraVert
     \left\langle{#1}\right\rangle\endgroup}
}
\def\BraVert{\@ifnextchar|{\|\@gobble}
     {\egroup\,\mid@vertical\,\bgroup}}
\def\BraDoubleVert{\egroup\,\mid@dblvertical\,\bgroup}
\let\SavedDoubleVert\relax

{\catcode`\|=\active
  \xdef\set{\protect\expandafter\noexpand\csname set \endcsname}
  \expandafter\gdef\csname set \endcsname#1{\mathinner
        {\lbrace\,{\mathcode`\|32768\let|\midvert #1}\,\rbrace}}
  \xdef\Set{\protect\expandafter\noexpand\csname Set \endcsname}
  \expandafter\gdef\csname Set \endcsname#1{\left\{%
     \ifx\SavedDoubleVert\relax \let\SavedDoubleVert\|\fi
     \:{\let\|\SetDoubleVert
     \mathcode`\|32768\let|\SetVert
     #1}\:\right\}}
}
\def\midvert{\egroup\mid\bgroup}
\def\SetVert{\@ifnextchar|{\|\@gobble}
    {\egroup\;\mid@vertical\;\bgroup}}
\def\SetDoubleVert{\egroup\;\mid@dblvertical\;\bgroup}

\begingroup
 \edef\@tempa{\meaning\middle}
 \edef\@tempb{\string\middle}
\expandafter \endgroup \ifx\@tempa\@tempb
 \def\mid@vertical{\middle|}
 \def\mid@dblvertical{\middle\SavedDoubleVert}
\else
 \def\mid@vertical{\mskip1mu\vrule\mskip1mu}
 \def\mid@dblvertical{\mskip1mu\vrule\mskip2.5mu\vrule\mskip1mu}
\fi

\usepackage{graphicx}
\usepackage[T1]{fontenc}
\usepackage[]{amsfonts}
\usepackage[]{amsmath}
\usepackage[]{rotating}
\usepackage[]{color}
\usepackage[]{float}
\usepackage[]{fancyhdr}
\usepackage[]{booktabs}
\usepackage[]{epsfig}
\usepackage{appendix}
\usepackage{amssymb}
\usepackage{psfrag}
\usepackage{setspace}

\begin{document}

\title{Conservation law of operator current in open quantum systems}

\author{J. Salmilehto}
\affiliation{Department of Applied Physics/COMP, Aalto University, P.O. Box 14100, FI-00076 AALTO, Finland}
\author{P. Solinas}
\affiliation{Department of Applied Physics/COMP, Aalto University, P.O. Box 14100, FI-00076 AALTO, Finland}
\affiliation{Low Temperature Laboratory, Aalto University, P.O. Box 13500, FI-00076 AALTO, Finland}
\author{M. M\"ott\"onen}
\affiliation{Department of Applied Physics/COMP, Aalto University, P.O. Box 14100, FI-00076 AALTO, Finland}
\affiliation{Low Temperature Laboratory, Aalto University, P.O. Box 13500, FI-00076 AALTO, Finland}


\begin{abstract}

We derive a fundamental conservation law of operator current for master equations describing reduced quantum systems. If this law is broken, the temporal integral of the current operator of an arbitrary system observable does not yield in general the change of that observable in the evolution. We study Lindblad-type master equations as examples and prove that the application of the secular approximation during their derivation results in a violation of the conservation law. We show that generally any violation of the law leads to artificial corrections to the complete quantum dynamics, thus questioning the accuracy of the particular master equation.

\end{abstract}

\maketitle

\section{Introduction}

Quantum master equations are a valuable tool when describing the dynamics of open systems. However, the typically employed reduced-density-operator theory does not \emph{a priori} guarantee that the resulting evolution maintains all necessary physical properties. A well-known example of the pursuit for these properties is given in the case of quantum Markov processes by the Lindblad form describing the most general generators of the quantum dynamical semigroup~\cite{cmp48/119, jmp17/821}. Even though this form and its time-dependent generalizations ensure certain critical properties of quantum evolution~\cite{cmp48/119, jmp17/821, tToOQS}, they do not account for time-local conservation of observables. Additionally, many microscopic derivations of master equations exploit the \textit{secular approximation} \cite{API} that has been shown to lead to non-physical behavior including nonconservation of electric charge~\cite{prl105/030401, prb82/134517, pe42/565, prb80/245107}. Related to the conservation, the continuity equation for current has been studied in the coarse-grained description of the reduced quantum dynamics in Refs.~\cite{prl93/160404, pra73/064101, AiCFTEnglman}.

In this paper, we introduce a general framework for the conservation law that all master equations for reduced quantum systems should ideally follow. The evolution obeying the law ensures that the temporal integral of the current operator of an arbitrary system observable, as obtained from the commutator with the Hamiltonian of the complete system, yields the change of that observable in time. In other words, the current flowing into the system equals the current obtained by it. 
As examples, we apply the conservation law to a few typical derivations of master equations leading to the Lindblad form and show that the secular approximation leads to nonconservation. 
Hence, Lindblad-type master equations do not intrinsically guarantee conservation for all observables.

\section{Conservation of operator current} 

Let us consider a quantum system described by a density operator $\hat{\rho}$. We differentiate a subsystem $S$ described by a reduced density operator $\hat{\rho}_S = \mathrm{Tr}_E \{ \hat{\rho} \}$, where the trace is over the remaining environmental degrees of freedom, and we denote a general $S$-observable as $\hat{G}$. We refer to the time derivate of the expectation value of the observable as \textit{operator current} and write it as
\begin{equation}
\frac{d}{dt}\braket{\hat{G}} = \mathrm{Tr} \left\{ \frac{d\hat{\rho}}{dt} \hat{G} \right\} + \mathrm{Tr} \left\{ \hat{\rho} \frac{d\hat{G}}{dt} \right\}.
\label{eq:op_current1}
\end{equation}
The von Neumann equation $\frac{d}{dt} \hat{\rho} = -\frac{i}{\hbar}[\hat{H},\hat{\rho}]$, results in the Ehrenfest theorem stating that \footnote{The results presented in this paper could as well be derived in the Heisenberg picture, in the case of which expectation values are not needed in the definition of the operator current. If $\hat{G}_H$ is the observable and $\hat{H}_H$ the total Hamiltonian in the Heisenberg picture, the current is described by the Heisenberg equation of motion $\frac{d}{dt}\hat{G}_H = -\frac{i}{\hbar} [\hat{G}_H,\hat{H}_H] + \frac{\partial \hat{G}_H}{\partial t}$, where $\partial / \partial t$ denotes a partial derivative with respect to the explicit time dependence of the observable in the Schr{\"o}dinger picture.}
\begin{equation}
\frac{d}{dt}\braket{\hat{G}} = -\frac{i}{\hbar} \mathrm{Tr} \{ \hat{\rho}[\hat{G},\hat{H}] \} + \mathrm{Tr} \left\{ \hat{\rho} \frac{d\hat{G}}{dt} \right\},
\label{eq:Ehrenfest}
\end{equation}
where $\hat{H}$ is the Hamiltonian of the total system. In order to relate this to the evolution of the subsystem of interest, we write the total Hamiltonian in the general form $\hat{H} = \hat{H}_S \otimes \hat{I}_E + \hat{I}_S \otimes \hat{H}_E + \hat{H}_I$, where we have separated Hamiltonians for the system, the environment, and the interaction between them, respectively. Using the full form of the Hamiltonian results in
\begin{equation}
\begin{split}
\frac{d}{dt}\braket{\hat{G}} &= -\frac{i}{\hbar} \left( \mathrm{Tr}_S \{ \hat{\rho}_S[\hat{G},\hat{H}_S] \} + \mathrm{Tr} \{ \hat{\rho}[\hat{G},\hat{H}_I] \} \right) \\ &+ \mathrm{Tr}_S \left\{ \hat{\rho}_S \frac{d\hat{G}}{dt} \right\},
\end{split}
\label{eq:op_current2}
\end{equation}
yielding our first definition for the operator current. We have denoted the trace over the subsystem degrees of freedom by $\mathrm{Tr}_S$. The current is comprised of three separate contributions. The first and third terms relate to the evolution of the closed system, and they are affected by the environment only through $\hat{\rho}_S$. The second term describes current induced by the interaction with the environment and vanishes for closed systems.

To illustrate how decoupling of the eigenstate populations and the coherence between them leads to nonphysical behavior, we provide a simple example. Consider a two-level system whose Hilbert space is $\mathcal{H}_S = \mathrm{span} ( \{ \ket{g},\ket{e} \})$, where $\hat{H}_S\ket{i} = E_i\ket{i}$ and inner products for an arbitrary system operator $\hat{O}_S$ are defined as $\braket{s|\hat{O}_S|p} = O_{sp}^S$, where $s,p \in \{g,e\}$. Assume that the system starts from a fully excited state $\tilde{\rho}_{ee}^S = 1$ and $\tilde{\rho}_{gg}^S = \tilde{\rho}_{ge}^S = 0$, and $\hat{H}_I = \hat{G} \otimes \hat{E}$, where $\hat{E}$ is any nontrivial environment operator and $\hat{G}$ is time independent. We assume that $\hat{G}$ is not diagonal in the eigenspace of the system Hamiltonian so that the system has a nonzero relaxation rate to the ground state. We consider a zero-temperature environment, and hence the system relaxes to the ground state and we have a stationary state $\bar{\rho}_{gg}^S = 1$ and $\bar{\rho}_{ee}^S = \bar{\rho}_{ge}^S = 0$. The expectation value of an observable assumes the general form $\braket{\hat{G}} = (G_{gg}-G_{ee})\rho_{gg}^S + 2 \Re e(\rho_{ge}^SG_{eg}) + G_{ee}$ so that, in the long-time limit, the temporal change in the expectation value becomes $\Delta \braket{\hat{G}} = G_{gg}-G_{ee}$, which is nonzero for an almost arbitrary operator $\hat{G}$. Equation~(\ref{eq:op_current2}) yields a current operator for $\hat{G}$ as $\hat{I}_G = -\frac{i}{\hbar} [\hat{G},\hat{H}_S]$ corresponding to the usual definition for subsystem current operators \cite{QM}. Hence, we have $\braket{\hat{I}_G} = -2 \omega_{01} \Im m(\rho_{ge}^S G_{eg})$, where $\omega_{01}=(E_e-E_g)/\hbar$, so that the integrated current becomes $\int \braket{\hat{I}_G} dt  = -2 \omega_{01} \Im m(G_{eg}\int \rho_{ge}^Sdt)$. Up to this point, the example has been on a very general level and no approximations on the dynamics have been invoked. However, if the populations and coherences decouple in the description of the dynamics for $\hat{\rho}_S$, our assumption of the initial state implies that $\rho_{ge}^S=0$ at all times. Thus $\braket{\hat{I}_G} = 0$ at all times yielding $\Delta \braket{\hat{G}} \neq \int \braket{\hat{I}_G}dt$ for an almost arbitrary $\hat{G}$. Hence, the local conservation of the operator current breaks down in the sense that the current cannot accurately describe the temporal change of the observable. In the following, we formulate a general condition ensuring this conservation.

Let the temporal evolution of the reduced system be described by a master equation as
\begin{equation}
\begin{split}
\frac{d}{dt}\hat{\rho}_S = -\frac{i}{\hbar} [\hat{H}_S,\hat{\rho}_S] + \hat{D},
\end{split}
\label{eq:master}
\end{equation}
where we have separated the part relating to unitary evolution from the generator and $\hat{D}=\hat{D}(\hat{\rho}_S,t)$ represents a generalized dissipator; that is, it also accounts for any unitary contribution stemming from the system--environment interaction. Combining Eqs.~(\ref{eq:op_current1}) and~(\ref{eq:master}) results in our second definition for the operator current:
\begin{equation}
\begin{split}
\frac{d}{dt}\braket{\hat{G}} &= -\frac{i}{\hbar} \mathrm{Tr}_S \{ \hat{\rho}_S[\hat{G},\hat{H}_S] \} + \mathrm{Tr}_S \{ \hat{D}\hat{G} \} \\ &+ \mathrm{Tr}_S \left\{ \hat{\rho}_S \frac{d\hat{G}}{dt} \right\}.
\end{split}
\label{eq:op_current3}
\end{equation}
Thus, we have two fundamental definitions provided by Eqs.~(\ref{eq:op_current2}) and (\ref{eq:op_current3}) leading to a necessary and sufficient condition for the conservation of the operator current:
\begin{equation}
\begin{split}
-\frac{i}{\hbar} \mathrm{Tr} \{ \hat{\rho} [\hat{G},\hat{H}_I] \} = \mathrm{Tr}_S \{ \hat{D}\hat{G} \}.
\end{split}
\label{eq:conservation_law}
\end{equation}
This condition states that the dissipative current obtained from the master equation must be equal to the dissipative current related to the interaction Hamiltonian and ensures the conservation of operator current which we define as $\Delta \braket{\hat{G}} = \int \braket{\hat{I}_G}dt$, where $\Delta \braket{\hat{G}}$ is the temporal change given by the master equation and Eq.~(\ref{eq:op_current2}) defines $\braket{\hat{I}_G} = \frac{d}{dt}\braket{\hat{G}}$. In practice, the complete dynamics of the total density operator can be unknown, and hence it is convenient to cast Eq. (6) into the form $\textrm{Tr}_S\{\hat{D}\hat{G}\}=0$ for all $[\hat{G},\hat{H}_I]=0$. This condition emphasizes the fact that physical quantities which are conserved by the interaction Hamiltonian, have to be conserved in the reduced dynamics; that is, the dissipative current must vanish in this case.

Note that the preceding derivation required that the master equations describe the system dynamics exactly. However, a typical derivation of a quantum master equation involves a set of approximations resulting in an approximate description of the dynamics. The conservation law is a valuable tool also in this case: We can take any master equation determining the reduced-system evolution and define a set of corresponding total quantum states $\{ |\Psi\rangle \}$ as the ones satisfying $\hat{\rho}_S = \mathrm{Tr}_E \{ |\Psi\rangle \langle\Psi| \}$~\footnote{At least one such state exists provided that the reduced density operator is positive semidefinite and Hermitian and its trace is unity and that the environment is high dimensional.}. For each of these states, the evolution is unitary, and hence we can define an operator corresponding to the total Hamiltonian $\hat{H}_A = \hat{H} + \hat{H}_{\delta}$. Thus the real approximate evolution, $\hat{\rho}_S$, corresponds to an exact evolution of a different system. As a consequence, Eq.~(\ref{eq:conservation_law}) yields $-\frac{i}{\hbar} \mathrm{Tr} \{ |\Psi\rangle \langle\Psi| [\hat{G},\hat{H}_I+\hat{H}_{\delta}] \} = \mathrm{Tr}_S \{ \hat{D}\hat{G} \}$. If the condition in Eq.~(\ref{eq:conservation_law}) is not obeyed naturally by the approximate master equation, we obtain $\hat{H}_{\delta} \neq 0$. Hence, an artificial effective Hamiltonian emerges in the complete description of the dynamics. Thus the conservation law provides an indicator of the reliability and accuracy of different approximations leading to reduced-system dynamics even if the complete quantum dynamics cannot be solved.

Let us return to the two-level example and apply the conservation law. We have $[\hat{G},\hat{H}_I]=0$ implying that the dissipative current vanishes. A general master equation yields $\mathrm{Tr}_S \{\hat{D}\hat{G} \} = (G_{gg}-G_{ee})D_{gg} + 2 \Re e\{ D_{ge}G_{eg} \}$, where we used $D_{gg}=-D_{ee}$ and $D_{ge} = D^*_{eg}$ stemming from the properties of the density operator through the master equation. If populations $\rho_{gg}^S$ and coherences $\rho_{ge}^S$ decouple as in typical master equation approaches, the conservation law only holds for constant populations, which is a contradiction. Hence, the accuracy of the approach is compromised as discussed above.

\section{Properties of dissipative current} 

Let us define the most general form for the interaction Hamiltonian as $\hat{H}_I = \sum_{\alpha} \hat{A}_{\alpha} \otimes \hat{B}_{\alpha}$ where $\hat{A}_{\alpha} = \hat{A}_{\alpha}^{\dagger}$ acts on the system degrees of freedom and $\hat{B}_{\alpha} = \hat{B}_{\alpha}^{\dagger}$ on the environment degrees of freedom. The dissipative current on the left-hand side of Eq.~(\ref{eq:conservation_law}) becomes
\begin{equation}
\begin{split}
-\frac{i}{\hbar} \mathrm{Tr} \{ \hat{\rho} [\hat{G},\hat{H}_I] \} = -\frac{i}{\hbar} \mathrm{Tr}_S \left\{ \sum_{\alpha} [\hat{A}_{\alpha},\mathrm{Tr}_E \{ \hat{B}_{\alpha} \hat{\rho} \} ] \hat{G} \right\},
\end{split}
\label{eq:diss_left1}
\end{equation}
allowing us to reduce the conservation law to a comparison of traces over $S$. Formulating operators $\mathrm{Tr}_E \{ \hat{B}_{\alpha} \hat{\rho} \}$ requires knowledge of the total system evolution and, hence, must be done for each system separately. However, an adequate condition for the disappearance of the dissipative current, not dependent on the time evolution, is evident: if $[\hat{G},\hat{H}_I] = \sum_{\alpha} [\hat{G},\hat{A}_{\alpha}] \otimes \hat{B}_{\alpha} = 0$, the dissipative current vanishes. The interaction Hamiltonian can always be expressed such that 
$\{ \hat{B}_{\alpha} \}$ forms an orthogonal basis of the environmental operator space, and hence the tensor product form implies that this condition is equivalent to $[\hat{G},\hat{A}_{\alpha}] = 0$ for each system operator in the decomposition.

A large range of microscopic derivations of master equations relies on the \textit{Born approximation} stating that the environment is only weakly coupled to the system. Thus, the density matrix of the environment is assumed to be negligibly affected by the interaction so that $\hat{\rho}(t) \approx \hat{\rho}_S(t) \otimes \hat{\rho}_E$. This results in $\mathrm{Tr}_E \{ \hat{B}_{\alpha} \hat{\rho} \} = \mathrm{Tr}_E \{ \hat{B}_{\alpha} \hat{\rho}_E \} \hat{\rho}_S(t) = \braket{\hat{B}_{\alpha}}_E \hat{\rho}_S(t)$ using $\braket{ \ }_E$ for the environment average. A noise source for which the environment average of the perturbation vanishes for each $\alpha$, an assumption used in a variety of derivations, leads apparently to a vanishing dissipative current on the right-hand side of Eq.~\eqref{eq:conservation_law}. This would naively imply that any derivation of the quantum master equation utilizing the Born approximation and the preceeding assumption should result in $\mathrm{Tr}_S \{\hat{D}\hat{G} \} = 0$. However, we will show that this does not generally apply and that the level, at which the approximation is performed, is the key. Performing it in the derivation of the master equation as usual allows for weak dissipative current, whereas performing it on the level of Eq.~\eqref{eq:op_current2} results in the artifact of total decoupling of the dissipative contribution.

\section{Lindblad form and secular approximation} 

To connect our general theory described above to a few important examples, we turn our attention to quantum Markov processes \cite{tToOQS} and study different microscopic derivations leading to master equations of the Lindblad form. The Lindblad form describes the most general form that the generator of a quantum dynamical semigroup can take, hence guaranteeing both the semigroup property and the properties of the dynamical map \cite{cmp48/119, jmp17/821}. However, the form itself is an abstract construction and does not imply operator current conservation. Hence, microscopic derivations leading to specific dissipators must be individually studied to see if they are in accordance with the conservation law. We are especially interested in derivations exploiting the secular approximation as it leads to the decoupling of populations and coherences, a feature which was shown above to result in nonphysical behavior in general.


\subsection{Singular-coupling limit}

Let us begin with the so-called singular-coupling limit, in which the coupling between the system and the environment is strong compared with the system Hamiltonian but weak compared with the bath Hamiltonian~\cite{tToOQS, prb80/033302}.
The master equation reads in the Schr\"odinger picture~\cite{tToOQS}
\begin{equation}
\begin{split}
\frac{d}{dt}\hat{\rho}_S = &- \frac{i}{\hbar}[\hat{H}_S+\hat{H}_{LS},\hat{\rho}_S] \\ &+ \sum_{\alpha\beta} \frac{\gamma_{\alpha\beta}}{2}([\hat{A}_{\beta},\hat{\rho}_S\hat{A}_{\alpha}] + [\hat{A}_{\beta}\hat{\rho}_S,\hat{A}_{\alpha}] ),
\end{split}
\label{eq:singular}
\end{equation}
which we have left in the so-called first standard form that can be explicitly transformed to the Lindblad form by a diagonalization of the Hermitian rate matrix $\{ \gamma_{\alpha\beta} \}$. The Lamb shift Hamiltonian is $\hat{H}_{LS} = \sum_{\alpha\beta} S_{\alpha\beta}\hat{A}_{\alpha}\hat{A}_{\beta}$. Note that $\gamma_{\alpha\beta}$ and $S_{\alpha\beta}$ are dependent on the Fourier transforms of the environment correlation functions. The only requirement for the correlation functions imposed by the derivation is sufficiently fast decay to accommodate the Markovian approximation.

Let us concentrate on the special case of vanishing dissipative current such that $[\hat{G},\hat{H}_I] = 0$. Since this implies $[\hat{A}_{\alpha},\hat{G}] = 0$ for each $\alpha$, it suffices to study $\hat{H}_I = \sum_{\alpha} \hat{A}_{\alpha} \otimes \hat{B}_{\alpha}=\hat{A} \otimes \hat{B}$. Using the generalized dissipator $\hat{D}_{\mathrm{sc}}$ from Eq.~(\ref{eq:singular}), we obtain
\begin{equation}
\begin{split}
\mathrm{Tr}_S \{ \hat{D}_{\mathrm{sc}}\hat{G} \} = &-\frac{iS}{\hbar} \mathrm{Tr}_S \{ \hat{A} \hat{\rho}_S [\hat{G},\hat{A}] \} \\ &+ \frac{\gamma}{2} \mathrm{Tr}_S \{ \hat{\rho}_S \hat{A} [\hat{G},\hat{A}]+\hat{A}\hat{\rho}_S [\hat{A},\hat{G}] \},
\end{split}
\label{eq:singular_diss}
\end{equation}
where $S$ and $\gamma$ are scalar constants. Above, we utilized the cyclicity of the trace. This expression vanishes due to the commutation of $\hat{A}$ and $\hat{G}$ and, hence, the operator current is conserved in the case of the vanishing dissipative current. We emphasize that even though the master equation was in the first standard form and utilized the Born--Markov approximation, the secular approximation was not used in its derivation.

\subsection{Weak-coupling limit}

Next, we study the derivation in the weak-coupling limit in which the secular approximation is necessary to achieve a Lindblad-type master equation. Again, it is sufficient to study interaction Hamiltonians of the form $\hat{H}_I=\hat{A} \otimes \hat{B}$. The master equation assumes in the Schr\"odinger picture the form
\begin{equation}
\begin{split}
&\frac{d}{dt}\hat{\rho}_S = -\frac{i}{\hbar}[\hat{H}_S+\hat{H}_{LS},\hat{\rho}_S] \\ &+ \sum_{\omega} \frac{\gamma({\omega})}{2} ([\hat{A}(\omega),\hat{\rho}_S\hat{A}^{\dagger}(\omega)]+[\hat{A}(\omega)\hat{\rho}_S,\hat{A}^{\dagger}(\omega)]),
\end{split}
\label{eq:weak}
\end{equation}
where $\hat{H}_{LS} = \sum_{\omega} S(\omega)\hat{A}^{\dagger}(\omega)\hat{A}(\omega)$. The eigenoperators are defined as $\hat{A}(\omega) = \sum_{\epsilon'-\epsilon = \hbar\omega} \hat{\Pi}(\epsilon)\hat{A}\hat{\Pi}(\epsilon')$, where $\hat{\Pi}$ are projections to the respective eigenspaces of $\hat{H}_S$ and the sum is over all eigenvalues $\epsilon$ and $\epsilon'$ with a fixed $\omega$. Note that the master equation is of the first standard form, and the parameters $\gamma(\omega)$ and $S(\omega)$ attain a dependence on the frequency difference $\omega$. We obtain
\begin{equation}
\begin{split}
& \mathrm{Tr}_S \{ \hat{D}_{\mathrm{wc}}\hat{G} \} = \sum_{\omega} \bigg( -\frac{i}{\hbar} S(\omega) \mathrm{Tr}_S \{ \hat{A}(\omega)\hat{\rho}_S[\hat{G},\hat{A}^{\dagger}(\omega)] \} \\ &+ \frac{\gamma(\omega)}{2} \mathrm{Tr}_S \{ \hat{\rho}_S\hat{A}^{\dagger}(\omega)[\hat{G},\hat{A}(\omega)] + \hat{A}(\omega)\hat{\rho}_S[\hat{A}^{\dagger}(\omega),\hat{G}] \} \bigg),
\end{split}
\label{eq:weak_diss}
\end{equation}
where $\hat{D}_{\mathrm{wc}}$ corresponds to the dissipator in Eq.~(\ref{eq:weak}). Assuming vanishing dissipative current due to commutation translates to $[\hat{A},\hat{G}] = \sum_{\omega}[\hat{A}(\omega),\hat{G}] = \sum_{\omega}[\hat{A}^{\dagger}(\omega),\hat{G}] = 0$, which does not necessarily result in a vanishing expression in Eq.~(\ref{eq:weak_diss}). However, if $\hat{G}$ commutes with all the eigenoperators individually, the operator current is conserved. One way to meet this special condition is to set $[\hat{G},\hat{\Pi}(\epsilon)] = 0$ for every $\epsilon$ implying that the observable $\hat{G}$ must be diagonal in the eigenbasis of $\hat{H}_S$ and hence cannot induce transitions. Again, this does not hold in general.

Comparison with the singular-coupling limit points to problems with the secular approximation. In order to determine if this is the cause of the nonconservation, we go to an earlier stage in the derivation of the master equation in the weak-coupling limit. Without the secular approximation, the Redfield-type master equation yields a dissipator $\hat{D}_{\mathrm{wc},I}^{\mathrm{nonsec}}$ in the interaction picture for which
\begin{equation}
\begin{split}
\mathrm{Tr}_S \{ \hat{D}_{\mathrm{wc},I}^{\mathrm{nonsec}}\hat{G}_I \} = &\sum_{\omega} \Gamma(\omega)e^{-i\omega t} \mathrm{Tr}_S \{ \hat{A}(\omega) \hat{\rho}_S \\ &\times \sum_{\omega'} e^{i\omega' t} [\hat{A}^{\dagger}(\omega'),\hat{G}_I] \} + \mathrm{c.c.},
\end{split}
\label{eq:weak_diss_nosec}
\end{equation}
where $\hat{G}_I = e^{i\hat{H}_S t} \hat{G} e^{-i\hat{H}_S t}$, $\Gamma(\omega)$ is a specific Fourier transform of the environment correlation functions and $\mathrm{c.c.}$ denotes a complex conjugate of the preceding term. Here, the construction of the eigenoperators yields $\sum_{\omega'} e^{i\omega' t} [\hat{A}^{\dagger}(\omega'),\hat{G}_I] = \sum_{\omega'} [e^{i\hat{H}_S t}\hat{A}^{\dagger}(\omega')e^{-i\hat{H}_S t},\hat{G}_I] = e^{i\hat{H}_S t} \sum_{\omega'} [\hat{A}^{\dagger}(\omega'),\hat{G}]e^{-i\hat{H}_S t} = e^{i\hat{H}_S t} [\hat{A},\hat{G}]e^{-i\hat{H}_S t} = 0$. Hence, we retrieve the operator current conservation for the vanishing dissipative current if the secular approximation is not performed.

\subsection{Weak-coupling limit for adiabatically driven systems}

In our last example, a time-dependent external field is used to drive a weakly coupled system adiabatically. See Refs.~\cite{prl105/030401, prb82/134517, pra82/062112, prb84/174507, prb83/214508, prb84/235140} for recent theoretical progress in this field. Using a superadiabatic master equation based on a perturbative expansion, it has been shown for two-level systems that the application of the secular approximation here results in nonconservation of the operator current~\cite{prb85/024527}. The current was found to be conserved if the secular approximation was dropped. To account for the exact effect of the steering, we approach the problem utilizing a modified Floquet mode basis~\cite{prb84/235140} where the master equation in the Schr\"odinger picture is given by
\begin{equation}
\begin{split}
\frac{d}{dt}\hat{\rho}_S = &-\frac{i}{\hbar}[\hat{H}_S+\hat{H}_{LS},\hat{\rho}_S] \\ &+ \frac{\gamma(0)}{2} ( [\hat{L}_0,\hat{\rho}_S\hat{L}^{\dagger}_0] + [\hat{L}_0\hat{\rho}_S,\hat{L}^{\dagger}_0] ) \\ &+ \sum_{\alpha \neq \beta} \frac{\gamma(\omega_{\alpha\beta})}{2} ( [\hat{L}_{\alpha\beta},\hat{\rho}_S\hat{L}^{\dagger}_{\alpha\beta}] + [\hat{L}_{\alpha\beta}\hat{\rho}_S,\hat{L}^{\dagger}_{\alpha\beta}] ),
\end{split}
\label{eq:floquet}
\end{equation}
where $\hat{H}_{LS} = \sum_{\alpha\beta} S(\omega_{\alpha\beta}) \hat{\Pi}(\beta) \hat{A} \hat{\Pi}(\alpha) \hat{A} \hat{\Pi}(\beta)$, $\hat{L}_0 = \sum_{\alpha} \hat{\Pi}(\alpha) \hat{A} \hat{\Pi}(\alpha)$, $\hat{L}_{\alpha\beta} = \hat{\Pi}(\alpha) \hat{A} \hat{\Pi}(\beta)$, and $\hat{\Pi}(x) = \ket{\phi_x(t)}\bra{\phi_x(t)}$ denotes a projection operator to the $x$th modified Floquet mode at time $t$. The parameters $\omega_{\alpha\beta}$ denote the angular frequencies when the modified modes are used, and the real-valued functions $\gamma(\omega_{\alpha\beta})$ and $S(\omega_{\alpha\beta})$ relate to certain Fourier transforms of the environment correlation function. Note that the rates and projection operators are time dependent as they describe dynamics in the Floquet basis. The generator in Eq.~(\ref{eq:floquet}) is of the Lindblad form at each time instant and is obtained by applying the secular approximation. The derivation is carried out for $\hat{H}_I = \hat{A} \otimes \hat{B}$ but we expect a similar result for a general decomposition. It turns out that Eq.~(\ref{eq:floquet}) does not necessarily result in vanishing dissipative current for $[\hat{A},\hat{G}] = 0$ and an arbitrary noise source. Similarly to the nondriven system, in the special case of $[\hat{G},\hat{\Pi}(\alpha)] = 0$ for every $\alpha$, the commutation leads to vanishing dissipative current. The difference in this special condition compared with the nondriven case is that instead of the observable being diagonal in the eigenspace of the system Hamiltonian, it needs to be diagonal in the Floquet basis at all times.

To clarify the role of the secular approximation, we can rewrite the master equation without applying it. In the interaction picture, the resulting Redfield-type dissipator $\hat{D}_{\mathrm{driven},I}^{\mathrm{nonsec}}$ gives
\begin{equation}
\begin{split}
&\mathrm{Tr}_S \{ \hat{D}_{\mathrm{driven},I}^{\mathrm{nonsec}} \hat{G}_I \} = \\  &\sum_{\alpha\alpha'} \Gamma(\omega_{\alpha\alpha'}) e^{-i \int_0^t dt' \omega_{\alpha\alpha'}} \mathrm{Tr}_S \{ \hat{U}^{\dagger}(\alpha)\hat{A}\hat{U}(\alpha') \hat{\rho}_S  \\ &\times \sum_{\beta\beta'} e^{i\int_0^t dt' \omega_{\beta\beta'}} [\hat{U}^{\dagger}(\beta')\hat{A}\hat{U}(\beta),\hat{G}_I] \} + \mathrm{c.c.},
\end{split}
\label{eq:floquet_diss_nonsec}
\end{equation}
where $\hat{G}_I$ denotes again the observable in the interaction picture and $\hat{U}(x) = \ket{\phi_x(t)}\bra{\phi_x(0)}$ denotes a propagator for the $x$th mode. Note that $\hat{A} = \sum_{\beta\beta'} \hat{\Pi}(\beta')\hat{A}\hat{\Pi}(\beta)$ so that in the interaction picture $\hat{A}_I = \sum_{\beta\beta'}e^{i\int_0^t dt' \omega_{\beta\beta'}} \hat{U}^{\dagger}(\beta')\hat{A}\hat{U}(\beta)$. Hence $\sum_{\beta\beta'} e^{i\int_0^t dt' \omega_{\beta\beta'}} [\hat{U}^{\dagger}(\beta')\hat{A}\hat{U}(\beta),\hat{G}_I] = [\hat{A}_I,\hat{G}_I] = 0$ since $[\hat{A},\hat{G}] = 0$. Thus, the dissipative current vanishes indicating conservation.

\section{Conclusions} 

We introduced a fundamental conservation law of operator current in open quantum systems ensuring that the current flowing into the system equals the current obtained by it. For example, different Lindblad-type master equations stemming from the secular approximation were found not to obey the law. In the future, our analysis provides a basic tool for exploring the regimes of validity of the different approximations employed in the reduced-density-operator theory for open quantum systems. The conservation law is crucial especially in cases where the operator current is of great interest.

\begin{acknowledgments}
We thank P. Jones for useful advice. We acknowledge the Academy of Finland, the V\"ais\"al\"a Foundation, the KAUTE Foundation and the Emil Aaltonen Foundation for financial support. We have received funding from the European Community's Seventh Framework Programme under Grant No. 238345 (GEOMDISS) and from the European Research Council under Grant No. 278117 (SINGLEOUT).
\end{acknowledgments}

\bibliography{localbib.bib}

\begin{thebibliography}{10}%
\makeatletter
\providecommand \@ifxundefined [1]{%
 \ifx #1\undefined \expandafter \@firstoftwo
 \else \expandafter \@secondoftwo
\fi
}%
\providecommand \@ifnum [1]{%
 \ifnum #1\expandafter \@firstoftwo
 \else \expandafter \@secondoftwo
\fi
}%
\providecommand \enquote [1]{``#1''}%
\providecommand \bibnamefont  [1]{#1}%
\providecommand \bibfnamefont [1]{#1}%
\providecommand \citenamefont [1]{#1}%
\providecommand\href[0]{\@sanitize\@href}%
\providecommand\@href[1]{\endgroup\@@startlink{#1}\endgroup\@@href}%
\providecommand\@@href[1]{#1\@@endlink}%
\providecommand \@sanitize [0]{\begingroup\catcode`\&12\catcode`\#12\relax}%
\@ifxundefined \pdfoutput {\@firstoftwo}{%
 \@ifnum{\z@=\pdfoutput}{\@firstoftwo}{\@secondoftwo}%
}{%
 \providecommand\@@startlink[1]{\leavevmode\special{html:<a href="#1">}}%
 \providecommand\@@endlink[0]{\special{html:</a>}}%
}{%
 \providecommand\@@startlink[1]{%
  \leavevmode
  \pdfstartlink
   attr{/Border[0 0 1 ]/H/I/C[0 1 1]}%
   user{/Subtype/Link/A<</Type/Action/S/URI/URI(#1)>>}%
  \relax
 }%
 \providecommand\@@endlink[0]{\pdfendlink}%
}%
\providecommand \url  [0]{\begingroup\@sanitize \@url }%
\providecommand \@url [1]{\endgroup\@href {#1}{\urlprefix}}%
\providecommand \urlprefix [0]{URL }%
\providecommand \Eprint[0]{\href }%
\@ifxundefined \urlstyle {%
  \providecommand \doi [1]{doi:\discretionary{}{}{}#1}%
}{%
  \providecommand \doi [0]{doi:\discretionary{}{}{}\begingroup
  \urlstyle{rm}\Url }%
}%
\providecommand \doibase [0]{http://dx.doi.org/}%
\providecommand \Doi[1]{\href{\doibase#1}}%
\providecommand \bibAnnote [3]{%
  \BibitemShut{#1}%
  \begin{quotation}\noindent
    \textsc{Key:}\ #2\\\textsc{Annotation:}\ #3%
  \end{quotation}%
}%
\providecommand \bibAnnoteFile [2]{%
  \IfFileExists{#2}{\bibAnnote {#1} {#2} {\input{#2}}}{}%
}%
\providecommand \typeout [0]{\immediate \write \m@ne }%
\providecommand \selectlanguage [0]{\@gobble}%
\providecommand \bibinfo [0]{\@secondoftwo}%
\providecommand \bibfield [0]{\@secondoftwo}%
\providecommand \translation [1]{[#1]}%
\providecommand \BibitemOpen[0]{}%
\providecommand \bibitemStop [0]{}%
\providecommand \bibitemNoStop [0]{.\EOS\space}%
\providecommand \EOS [0]{\spacefactor3000\relax}%
\providecommand \BibitemShut [1]{\csname bibitem#1\endcsname}%
\bibitem{cmp48/119}%
  \BibitemOpen
  \bibfield{author}{%
  \bibinfo {author} {\bibfnamefont{G.}~\bibnamefont{Lindblad}},\ }%
  \bibfield{journal}{%
  \bibinfo {journal} {Commun. Math. Phys.}\ }%
  \textbf{\bibinfo {volume} {48}},\ \bibinfo {pages} {119} (\bibinfo {year}
  {1976})%
  \bibAnnoteFile{NoStop}{cmp48/119}%
\bibitem{jmp17/821}%
  \BibitemOpen
  \bibfield{author}{%
  \bibinfo {author} {\bibfnamefont{V.}~\bibnamefont{Gorini}}, \bibinfo {author}
  {\bibfnamefont{A.}~\bibnamefont{Kossakowski}},\ and\ \bibinfo {author}
  {\bibfnamefont{E.~C.~G.}\ \bibnamefont{Sudarshan}},\ }%
  \bibfield{journal}{%
  \bibinfo {journal} {J. Math. Phys.}\ }%
  \textbf{\bibinfo {volume} {17}},\ \bibinfo {pages} {821} (\bibinfo {year}
  {1976})%
  \bibAnnoteFile{NoStop}{jmp17/821}%
\bibitem{tToOQS}%
  \BibitemOpen
  \bibfield{author}{%
  \bibinfo {author} {\bibfnamefont{H.-P.}\ \bibnamefont{Breuer}}\ and\ \bibinfo
  {author} {\bibfnamefont{F.}~\bibnamefont{Pettrucione}},\ }%
  \emph{\bibinfo {title} {The Theory of Open Quantum Systems}}\ (\bibinfo
  {publisher} {Oxford University Press},\ \bibinfo {address} {Oxford},\
  \bibinfo {year} {2002})%
  \bibAnnoteFile{NoStop}{tToOQS}%
\bibitem{API}%
  \BibitemOpen
  \bibfield{author}{%
  \bibinfo {author} {\bibfnamefont{C.}~\bibnamefont{Cohen-Tannoudji}}, \bibinfo
  {author} {\bibfnamefont{J.}~\bibnamefont{Dupont-Roc}},\ and\ \bibinfo
  {author} {\bibfnamefont{G.}~\bibnamefont{Grynberg}},\ }%
  \emph{\bibinfo {title} {Atom-Photon Interactions}}\ (\bibinfo {publisher}
  {Wiley},\ \bibinfo {address} {New York},\ \bibinfo {year} {1992})%
  \bibAnnoteFile{NoStop}{API}%
\bibitem{prl105/030401}%
  \BibitemOpen
  \bibfield{author}{%
  \bibinfo {author} {\bibfnamefont{J.~P.}\ \bibnamefont{Pekola}}, \bibinfo
  {author} {\bibfnamefont{V.}~\bibnamefont{Brosco}}, \bibinfo {author}
  {\bibfnamefont{M.}~\bibnamefont{M\"ott\"onen}}, \bibinfo {author}
  {\bibfnamefont{P.}~\bibnamefont{Solinas}},\ and\ \bibinfo {author}
  {\bibfnamefont{A.}~\bibnamefont{Shnirman}},\ }%
  \bibfield{journal}{%
  \bibinfo {journal} {Phys. Rev. Lett.}\ }%
  \textbf{\bibinfo {volume} {105}},\ \bibinfo {pages} {030401} (\bibinfo {year}
  {2010})%
  \bibAnnoteFile{NoStop}{prl105/030401}%
\bibitem{prb82/134517}%
  \BibitemOpen
  \bibfield{author}{%
  \bibinfo {author} {\bibfnamefont{P.}~\bibnamefont{Solinas}}, \bibinfo
  {author} {\bibfnamefont{M.}~\bibnamefont{M\"ott\"onen}}, \bibinfo {author}
  {\bibfnamefont{J.}~\bibnamefont{Salmilehto}},\ and\ \bibinfo {author}
  {\bibfnamefont{J.~P.}\ \bibnamefont{Pekola}},\ }%
  \bibfield{journal}{%
  \bibinfo {journal} {Phys. Rev. B}\ }%
  \textbf{\bibinfo {volume} {82}},\ \bibinfo {pages} {134517} (\bibinfo {year}
  {2010})%
  \bibAnnoteFile{NoStop}{prb82/134517}%
\bibitem{pe42/565}%
  \BibitemOpen
  \bibfield{author}{%
  \bibinfo {author} {\bibfnamefont{J.}~\bibnamefont{Pracha\v{r}}}\ and\
  \bibinfo {author} {\bibfnamefont{T.}~\bibnamefont{Novotn\'y}},\ }%
  \bibfield{journal}{%
  \bibinfo {journal} {Physica E}\ }%
  \textbf{\bibinfo {volume} {42}},\ \bibinfo {pages} {565} (\bibinfo {year}
  {2010})%
  \bibAnnoteFile{NoStop}{pe42/565}%
\bibitem{prb80/245107}%
  \BibitemOpen
  \bibfield{author}{%
  \bibinfo {author} {\bibfnamefont{G.}~\bibnamefont{Schaller}}, \bibinfo
  {author} {\bibfnamefont{G.}~\bibnamefont{Kie{\ss}lich}},\ and\ \bibinfo
  {author} {\bibfnamefont{T.}~\bibnamefont{Brandes}},\ }%
  \bibfield{journal}{%
  \bibinfo {journal} {Phys. Rev. B}\ }%
  \textbf{\bibinfo {volume} {80}},\ \bibinfo {pages} {245107} (\bibinfo {year}
  {2009})%
  \bibAnnoteFile{NoStop}{prb80/245107}%
\bibitem{prl93/160404}%
  \BibitemOpen
  \bibfield{author}{%
  \bibinfo {author} {\bibfnamefont{R.}~\bibnamefont{Gebauer}}\ and\ \bibinfo
  {author} {\bibfnamefont{R.}~\bibnamefont{Car}},\ }%
  \bibfield{journal}{%
  \bibinfo {journal} {Phys. Rev. Lett.}\ }%
  \textbf{\bibinfo {volume} {93}},\ \bibinfo {pages} {160404} (\bibinfo {year}
  {2004})%
  \bibAnnoteFile{NoStop}{prl93/160404}%
\bibitem{pra73/064101}%
  \BibitemOpen
  \bibfield{author}{%
  \bibinfo {author} {\bibfnamefont{A.}~\bibnamefont{Bodor}}\ and\ \bibinfo
  {author} {\bibfnamefont{L.}~\bibnamefont{Di\'osi}},\ }%
  \bibfield{journal}{%
  \bibinfo {journal} {Phys. Rev. A}\ }%
  \textbf{\bibinfo {volume} {73}},\ \bibinfo {pages} {064101} (\bibinfo {year}
  {2006})%
  \bibAnnoteFile{NoStop}{pra73/064101}%
\bibitem{AiCFTEnglman}%
  \BibitemOpen
  \bibfield{author}{%
  \bibinfo {author} {\bibfnamefont{R.}~\bibnamefont{Englman}}\ and\ \bibinfo
  {author} {\bibfnamefont{A.}~\bibnamefont{Yahalom}},\ }%
  \emph{\bibinfo {title} {\textnormal{in} Advances in Classical Field Theory}}\
  (\bibinfo {publisher} {Bentham},\ \bibinfo {address} {Sharjah},\ \bibinfo
  {year} {2011})%
  \bibAnnoteFile{NoStop}{AiCFTEnglman}%
\bibitem{Note1}%
  \BibitemOpen
  \bibinfo {note} {The results presented in this paper could as well be derived
  in the Heisenberg picture, in the case of which expectation values are not
  needed in the definition of the operator current. If $\protect \mathaccentV
  {hat}05E{G}_H$ is the observable and $\protect \mathaccentV {hat}05E{H}_H$
  the total Hamiltonian in the Heisenberg picture, the current is described by
  the Heisenberg equation of motion $\protect \frac {d}{dt}\protect
  \mathaccentV {hat}05E{G}_H = -\protect \frac {i}{\hbar } [\protect
  \mathaccentV {hat}05E{G}_H,\protect \mathaccentV {hat}05E{H}_H] + \protect
  \frac {\partial \protect \mathaccentV {hat}05E{G}_H}{\partial t}$, where
  $\partial / \partial t$ denotes a partial derivative with respect to the
  explicit time dependence of the observable in the Schr{\"o}dinger picture.}%
  \bibAnnoteFile{Stop}{Note1}%
\bibitem{QM}%
  \BibitemOpen
  \bibfield{author}{%
  \bibinfo {author} {\bibfnamefont{F.}~\bibnamefont{Schwabl}},\ }%
  \emph{\bibinfo {title} {Quantum Mechanics (4th Edition)}}\ (\bibinfo
  {publisher} {Springer},\ \bibinfo {address} {New York},\ \bibinfo {year}
  {2007})%
  \bibAnnoteFile{NoStop}{QM}%
\bibitem{Note2}%
  \BibitemOpen
  \bibinfo {note} {At least one such state exists provided that the reduced
  density operator is positive semidefinite and Hermitian and its trace is
  unity and that the environment is high dimensional.}%
  \bibAnnoteFile{Stop}{Note2}%
\bibitem{prb80/033302}%
  \BibitemOpen
  \bibfield{author}{%
  \bibinfo {author} {\bibfnamefont{M.~G.}\ \bibnamefont{Schultz}}\ and\
  \bibinfo {author} {\bibfnamefont{F.}~\bibnamefont{von Oppen}},\ }%
  \bibfield{journal}{%
  \bibinfo {journal} {Phys. Rev. B}\ }%
  \textbf{\bibinfo {volume} {80}},\ \bibinfo {pages} {033302} (\bibinfo {year}
  {2009})%
  \bibAnnoteFile{NoStop}{prb80/033302}%
\bibitem{pra82/062112}%
  \BibitemOpen
  \bibfield{author}{%
  \bibinfo {author} {\bibfnamefont{J.}~\bibnamefont{Salmilehto}}, \bibinfo
  {author} {\bibfnamefont{P.}~\bibnamefont{Solinas}}, \bibinfo {author}
  {\bibfnamefont{J.}~\bibnamefont{Ankerhold}},\ and\ \bibinfo {author}
  {\bibfnamefont{M.}~\bibnamefont{M\"ott\"onen}},\ }%
  \bibfield{journal}{%
  \bibinfo {journal} {Phys. Rev. A}\ }%
  \textbf{\bibinfo {volume} {82}},\ \bibinfo {pages} {062112} (\bibinfo {year}
  {2010})%
  \bibAnnoteFile{NoStop}{pra82/062112}%
\bibitem{prb84/174507}%
  \BibitemOpen
  \bibfield{author}{%
  \bibinfo {author} {\bibfnamefont{J.}~\bibnamefont{Salmilehto}}\ and\ \bibinfo
  {author} {\bibfnamefont{M.}~\bibnamefont{M\"ott\"onen}},\ }%
  \bibfield{journal}{%
  \bibinfo {journal} {Phys. Rev. B}\ }%
  \textbf{\bibinfo {volume} {84}},\ \bibinfo {pages} {174507} (\bibinfo {year}
  {2011})%
  \bibAnnoteFile{NoStop}{prb84/174507}%
\bibitem{prb83/214508}%
  \BibitemOpen
  \bibfield{author}{%
  \bibinfo {author} {\bibfnamefont{A.}~\bibnamefont{Russomanno}}, \bibinfo
  {author} {\bibfnamefont{S.}~\bibnamefont{Pugnetti}}, \bibinfo {author}
  {\bibfnamefont{V.}~\bibnamefont{Brosco}},\ and\ \bibinfo {author}
  {\bibfnamefont{R.}~\bibnamefont{Fazio}},\ }%
  \bibfield{journal}{%
  \bibinfo {journal} {Phys. Rev. B}\ }%
  \textbf{\bibinfo {volume} {83}},\ \bibinfo {pages} {214508} (\bibinfo {year}
  {2011})%
  \bibAnnoteFile{NoStop}{prb83/214508}%
\bibitem{prb84/235140}%
  \BibitemOpen
  \bibfield{author}{%
  \bibinfo {author} {\bibfnamefont{I.}~\bibnamefont{Kamleitner}}\ and\ \bibinfo
  {author} {\bibfnamefont{A.}~\bibnamefont{Shnirman}},\ }%
  \bibfield{journal}{%
  \bibinfo {journal} {Phys. Rev. B}\ }%
  \textbf{\bibinfo {volume} {84}},\ \bibinfo {pages} {235140} (\bibinfo {year}
  {2011})%
  \bibAnnoteFile{NoStop}{prb84/235140}%
\bibitem{prb85/024527}%
  \BibitemOpen
  \bibfield{author}{%
  \bibinfo {author} {\bibfnamefont{P.}~\bibnamefont{Solinas}}, \bibinfo
  {author} {\bibfnamefont{M.}~\bibnamefont{M\"ott\"onen}}, \bibinfo {author}
  {\bibfnamefont{J.}~\bibnamefont{Salmilehto}},\ and\ \bibinfo {author}
  {\bibfnamefont{J.~P.}\ \bibnamefont{Pekola}},\ }%
  \bibfield{journal}{%
  \bibinfo {journal} {Phys. Rev. B}\ }%
  \textbf{\bibinfo {volume} {85}},\ \bibinfo {pages} {024527} (\bibinfo {year}
  {2012})%
  \bibAnnoteFile{NoStop}{prb85/024527}%
\end{thebibliography}%

\end{document}